\documentclass[12pt]{article}
\usepackage{epsfig}

\def\beq{\begin{equation}}
\def\eeq#1{\label{#1}\end{equation}}
\def\eeqn{\end{equation}}

\def\beqa{\begin{eqnarray}}
\def\eeqa#1{\label{#1}\end{eqnarray}}
\def\eeqan{\end{eqnarray}}

\let\bar=\overbar

\def\Dslash{\not{\hbox{\kern-4pt $D$}}}
\def\dslash{\not{\hbox{\kern-2pt $\del$}}}

\def\msb{{\bar{\ssstyle M \kern -1pt S}}}

\def\Title#1{\begin{center} {\Large {\bf #1} } \end{center}}

\begin{document}

\Title{X-ray lines, progenitor and beaming of Gamma-Ray Bursts}

\bigskip\bigskip


\begin{raggedright}  

{\it Davide Lazzati\index{Lazzati, D.}\\
Institute of Astronomy\\
University of Cambridge\\
CB3 0HA Cambridge, U.K.}
\bigskip\bigskip
\end{raggedright}

\begin{abstract}
X-ray emission features have been detected in at least five GRB
afterglows. Their detection and properties are of great importance in
our understanding of the nature of the GRB progenitor. Moreover, they
provide us with an un-collimated source of photons, allowing us to put
firm constraints on the total energy produced in GRBs, and therefore
on their degree of collimation. I will review the radiation mechanisms
and geometries proposed as sources of the line emission, and discuss
the energy budget of the lines and their implication for the overall
energetics of the phenomenon.
\end{abstract}

\section{Introduction}

The detection of X-ray emission lines in the afterglow of Gamma-Ray
Bursts (GRBs) allows us to probe the close GRB environment and
therefore gives clues on the nature of the progenitor. Moreover, being
un-collimated, line emission can be used to constrain the energy budget
of the bursts and therefore their beaming angle. 

There are at least five GRBs to date with evidence for X-ray emission
features in their early X-ray afterglow. Two of them were detected by
BeppoSAX (GRB~970508~\cite{Pi99} and GRB~000214~\cite{An00}), one by
ASCA (GRB~970828~\cite{Yo01}), one by Chandra (GRB~991216~\cite{Pi00})
and one by XMM-Newton (GRB~011211~\cite{Re02}). Even though none of
these features has a strictly compelling evidence (all are less than
$5\sigma$ statistical confidence), their number, the coincidence with
expected emission lines (in all the cases in which an alternative
redshift is known) and the fact that the features have been detected
by all the X-ray missions capable of doing so, make the presence of
these features difficult to challenge. The properties of the detected
lines are summarized in Table~\ref{tab1}.  Additional evidence for the
presence of lines or line complexes were also suggested for two more
XMM-Newton spectra of GRB~001025A and GRB~010220~\cite{Wa02} and the
detection of a transient absorption feature in the X-ray prompt
emission of GRB~990705~\cite{Am00}.  Finally it should be mentioned
that there are cases in which, despite prompt and deep searches, the
presence of emission features could be excluded at flux levels and
equivalent widths (EW) much fainter than the detected ones (e.g. in
GRB~020405~\cite{Mi02}).

\begin{table}[h!]
\centerline{
\begin{tabular}{l|cccc}
GRB               & 970508   & 970828      & 991216     & 000214      \\ 
\hline \hline
$z_{\rm{opt}}$     & 0.835     & 0.958$^\ast$ & 1.02      & 0.46$^\ast$    \\
Line ID          & Fe        &  Fe          & Fe        & Fe          \\
$F_{\rm{line}}^{[a]}$  & $30\pm10$  & $15\pm8$     & $17\pm5$    & $6.7\pm2.2$  \\
$h\nu$ (keV)      & $3.4\pm0.3$ & $5\pm0.25$   & $3.5\pm0.06$ & $4.7\pm0.2$  \\
EW(keV) & 1.5   &  3  & 0.5   & 2 \\
$t_s-t_e^\ddag$    & 6--16      & 32--38      & 37--40      & 12--41      \\
$L^{\rm{iso}}_{44}$ (erg/s) & $12\pm4$   & $8.1\pm4.3$  & $11\pm3$     & $0.6\pm0.2$  \\
$E^{\rm{iso}}_{49}$ (erg)   & $3.6\pm1.3$ & $5\pm3$     & $7.7\pm2.3$  & $0.6\pm0.3$  \\
Ref.              & \cite{Pi99} & \cite{Yo01} & \cite{Pi00} & \cite{An00}  \\
$\theta_j^{[b]}$  & 16.7 & 4.1 & 2.9 & \\
$E_{\gamma,51}^{\rm{F01}[c]}$ & 0.23 & 0.57 & 0.69 & \\
$E_{\gamma,51}^{[d]}$ & 11--18 & 4.5--25 & 3--38 & 2--3 
\end{tabular}}

\bigskip
\bigskip

\centerline{
\begin{tabular}{l|cccc}
GRB               & 011211        & 011211      & 011211      & 011211 \\ 
\hline \hline
$z_{\rm{opt}}$     & 2.14         & 2.14         & 2.14        & 2.14   \\
Line ID          & Si            & S           & Ar          & Tot$^{\dag}$ \\
$F_{\rm{line}}^{[a]}$  & $1.1\pm0.3$   & $1.\pm0.3$    & $0.7\pm0.3$  & $4\pm1.6$ \\
$h\nu$ (keV)      & $0.71\pm0.02$ & $0.88\pm0.01$ & $1.21\pm0.02$ & - \\
EW (keV)          & 0.43 & 0.48 & 0.46 & \\
$t_s-t_e^\ddag$    & 11-12.4     & 11-12.4      & 11-12.4      & 11-12.4 \\
$L^{\rm{iso}}_{44}$ (erg/s) & $4.3\pm1.2$  & $3.9\pm1.1$   & $2.6\pm1$     & $15.6\pm6.3$ \\
$E^{\rm{iso}}_{49}$ (erg)   & $0.6\pm0.2$  & $0.55\pm0.16$ & $0.4\pm0.14$ & $2.2\pm0.9$ \\
Ref.              & \cite{Re02} & \cite{Re02} & \cite{Re02} & \cite{Re02} \\
$E_{\gamma,51}^{[d]}$ &&&& 0.5--4.4 
\end{tabular}}
\caption{{Properties of the lines detected so far in the early afterglows of GRBs.
~~$^\ast$:~these bursts do not have an optical determination of the
redshift, which is inferred from the X-ray line, identified as a
FeXXVI $K_\alpha$ line. ~~$^\dag$~In this column all the five lines
detected by Reeves et al.~\cite{Re02} are added
together. ~~$^\ddag$~the start and end times of the observations, or
the time at which the line is observed to disappear. ~~$^{[a]}$:~units
of erg~cm$^{-2}$~s$^{-1}$. ~~$^{[b]}$:~the opening angle of the jet in
degrees as measured from the afterglow break
time~\cite{Fr01}. ~~$^{[c]}$:~total energy in $\gamma$-rays in units
of $10^{51}$~erg as measured by correcting the GRB fluence for the
solid angle~\cite{Fr01}. ~~$^{[d]}$:~total energy in $\gamma$-rays in
units of $10^{51}$~erg as measured from the emission lines
fluence~\cite{Gh02}. }
\label{tab1}}
\end{table}

\section{Line properties}

In this section I will summarize the ``average'' properties of the
detected features. I will in particular emphasize what are the
characteristics of the lines which are most difficult to explain in
the simplest GRB scenario. The measured properties of the individual
emission features are reported in Tab.~\ref{tab1}.

\paragraph{Luminosity}

The luminosity of the detected lines are large, and reproducing the
observed luminosities of these features is one of the big challenges
of theoretical models.
\begin{equation}
L\sim \mathrm{few}\times10^{44} \quad \mathrm{erg~s}^{-1}
\label{eq:lum}
\end{equation}

\paragraph{Equivalent width}
In addition to the large measured luminosities, the contrast of the
emission features with respect to the continuum is large. Again, this
is a challenging property for the models.
\begin{equation}
EW \sim 1 \quad \mathrm{keV}
\label{eq:ew}
\end{equation}

\paragraph{Frequency}
The observed frequency of the line emission is, in all cases for which
an optical redshift is known, consistent with emission from iron
(either neutral or highly ionized). In one case
(GRB~011211~\cite{Re02}) several lighter element lines were detected.
A proper velocity of the emitting material has been inferred in all
the ``good quality'' detections~\cite{Pi00,Re02}.

\paragraph{Duration}

The duration of the line emission is not completely clear, due to the
limitation of the performed observations. Most of the lines have been
observed to be active at comoving times
\begin{equation}
T\sim10\quad\mathrm{h}
\label{eq:dur}
\end{equation}
Some lines are observed to disappear during the
observations~\cite{Pi99,Yo01,Re02}, some are observed to remain
constant while the continuum fades~\cite{An00} while in some cases the
observations were too short to allow for any conclusion~\cite{Pi00}.

\section{First Consequences}

In this section I will review the first consequences that stem from
the mere detection of these lines with the fiducial properties
reported in the above section.

\paragraph{Reprocessing}

The lines are detected at a frequency consistent with the expected
emission lines from heavy elements in the rest frame of the host
galaxy of the bursts. At the time of detection, the fireball is
expected to be still highly relativistic~\cite{La99} ($\Gamma>10$).
Therefore, the fact that no blueshift (or only moderate~\cite{Re02})
is observed implies that the line photons are not produced directly
within the fireball itself, but by material at rest in the host
galaxy\footnote{In the framework of the cannonball model~\cite{Da02}, a
different identification for the lines is obtained. In this case, the
lines should be highly blueshifted hydrogen lines, produced by the
cannonballs themselves.}

\paragraph{Large densities}

It is easy to show that, given the observed properties of the lines, a
large density of the producing (or reprocessing) material is
required~\cite{La99}. Let us assume, in fact, that the lines are
produced in a low density medium, so that no recombination takes
place. Let us also consider iron lines (analogous conclusions can be
obtained for different elements). Given the line luminosity and
duration of the emission, we can compute the total number of line
photons emitted:
\begin{equation}
N_{\mathrm{Fe}}={{L_\mathrm{line}\,t_\mathrm{line}}\over
{h\nu_\mathrm{line}}} \sim 10^{57}
\label{eq:n1}
\end{equation}
where we used the luminosity and duration given in Eq.~\ref{eq:lum}
and~\ref{eq:dur}. Since each iron atom can produce at most 10 line
photons~\cite{La01} (given the effects of Auger auto-ionization) the
total mass of reprocessing iron can be computed and, assuming a solar
metallicity, the total mass of the reprocessing medium involved in the
line emission:
\begin{equation}
M={{N_\mathrm{Fe}}\over{10}} {{m_p}\over{4.67\times10^{-5}}} \sim 2000\,
 M_\odot
\end{equation}
The size of the emitting region can be constrained by considering
that the light travel time cannot be too long, otherwise the total
energy carried by line photons would be larger than the burst
energy. This yields a limit
\begin{equation}
R\le c\,t_\mathrm{line} \le 3\times10^{17} \quad\mathrm{cm}
\end{equation}
where we have allowed for a rather long duration of the line
$t_\mathrm{line}\sim100$~days. The lower limit on the density can be
now computed by simply dividing the mass over the volume occupied by
the medium:
\begin{equation}
n \ge {{M}\over{m_p\,V}} \sim {{N_\mathrm{Fe}}\over{10}}
{{1}\over{4.67\times10^{-5}}} {{3}\over{4\pi\,R^3}} \sim
2\times10^7 \quad\mathrm{cm}^{-3}
\label{eq:n}
\end{equation}
Analogous limits are obtained if, instead of assuming no
recombination, we ask a density large enough to allow for efficient
recombination of free electrons onto iron ions in the time scale of the
GRB duration.

\begin{figure}[htb]
\begin{center}
\epsfig{file=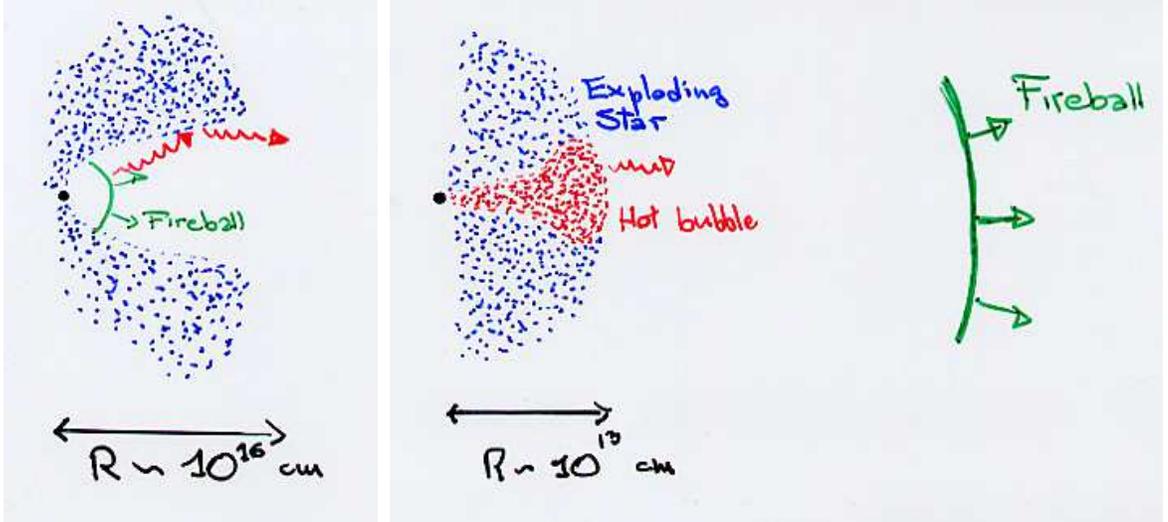,width=\textwidth}
\caption{{Cartoons showing the basic features of the two classes of 
models described in the text. The left panel shows the ``Geometry
Dominated'' (GD) models, while the right panel shows the ``Engine
Dominated'' (ED) models.}
\label{fig:car}}
\end{center}
\end{figure}

\section{The two classes of models}

The constraint on the density derived above poses a serious problem to
the overall modelling of the events. Consider, as an example, the case
of GRB~991216.Its afterglow lasted one month~\cite{Ha00}. A fireball
expanding in a high density medium should show a non-relativistic
transition which was not observed in the data. Moreover, all the
multiwavelength afterglow lightcurves can be modeled with low density
environments~\cite{PK02}.  Solving this apparent riddle of requiring a
very high density to produce the lines and a very low one to account
for the power-law afterglows is the most formidable tasks of all the
theories willing to explain how the lines have been formed.

The solutions that have been put forward to date can be divided in two
classes: those requiring a strong anisotropy of the surrounding medium
(a low density line of sight surrounded by high density material) and
those calling for a very small and dense reprocessor, which lies
inside the fireball and is surrounded by a low density ambient (see
Fig.~\ref{fig:car}). In these two classes of models, an important
difference is the reason for the observed duration of the line
emission.

\subsection{Geometry Dominated Models}

For historical reasons I will discuss first the class of Geometry
Dominated (GD) models, first discussed by Lazzati et al. in
1999~\cite{La99}. The basic idea of these class of models is that the
line emission comes from an extended region surrounding the GRB
progenitor. While the line emission takes place in a short period of
time (maybe the $\sim 100$ seconds of the burst duration only) the
observer at infinity receives line photons for a longer time, set by
the light crossing time of the emitting region. The typical size of
the emitting region is of the order of $R\sim10^{16}$ cm. Since the
region is large, the energy input is taken from the GRB and early
afterglow continuum, which is reprocessed through reflection off a
cold slab into line photons~\cite{Ba02,La02}.

The origin of the reprocessing material, which is likely to be
moderately enriched in heavy elements, is thought to be the remnant of
a recently exploded supernova~\cite{VS98} (SN) or a relic disk due to
the interaction of the burst progenitor with a companion
star~\cite{B00}.  The shape of this remnant was initially supposed to
be a partial covering shell, but the detection of outward velocities
of the reprocessing material made it more likely to think of a
funnel-like structure~\cite{Vi01} (see the left panel of
Fig.~\ref{fig:car}). Such an extreme geometry may be due to the
hyper-Eddington emission phase of the burst progenitor between the SN
and the burst proper explosions.

Even though the line emission mechanism is supposed to be the
reflection of the incident continuum off the cold dense material of
the funnel walls, a possible alternative is represented by an
optically thin shell of plasma, heated to $T\sim10^8$~K by the burst
photons themselves~\cite{Re02,L02}. Such an alternative, however,
would require an extremely clumped shell~\cite{La02}, and would be
quite difficult to explain the lack of detection of an iron (or cobalt
or nickel) line in the afterglow of GRB~011211~\cite{Re02}.

The existence of dense material in the surroundings of at least some
GRBs is corroborated by the detection of a transient absorption
feature in the afterglow of GRB~990705~\cite{Am00}. Such a feature can
be accounted for by the presence of a fraction of a solar mass of iron
at a distance $R\sim2\times10^{16}$~cm from the burst~\cite{La01}.

\subsection{Engine Dominated Models}
The class of GD models, even though provide us with a self-consistent
scenario for the production of the detected lines, requires a two step
explosion (first the SN and then the burst) and a rather extreme
geometrical setup. More recently, alternative models have been
presented. In this class of Engine Dominated models (ED), the line
reprocessing material is supposed to be contained in a very small
region, which is left behind by the fireball which therefore expands
in a low density medium, as required by the afterglow observations and
modelling (right panel of Fig.~\ref{fig:car}). The origin of this
dense material is the progenitor star itself~\cite{RM00,MR01}.

The open problem for ED models is to find a suitable source of energy
in order to power the line emission, since the fireball has already
escaped and cannot contribute to the line production. Two
possibilities have been discussed. In the first case, the energy may
be provided by the inner engine itself, which instead of turning off
abruptly at the end of the GRB emission remains active at a lower
level for a longer time, powering the line emission through
reflection off the surface of the funnel through which the fireball
crossed the star~\cite{RM00}.  Alternatively, the energy may be stored
in the cocoon that surrounds the jet as it crosses the
star~\cite{MR01}. In this case, however, it is difficult to have a
mixing of the jet and mantle material tuned to make the bubble hot
enough to power the line but not to explode as a second slower
fireball.

An additional problem of the ED models is the nickel decay time. In
usual SN explosions, in fact, iron is directly synthesized in very
small quantities, while unstable nickel is produced in large
quantities (up to a fraction of a solar mass). The unstable $^{56}$Ni
decays into $^{56}$Co in a time scale of $\sim6$ days which decays into
$^{56}$Fe after $\sim80$ days. In conclusion, in a typical young
supernova remnant, iron is the dominant element (over Ni and Co)
starting from $\sim100$ days after the explosion. For this reason, in
ED models, nickel emission lines should be observed rather then iron
ones. This problem can be ameliorated by downscattering~\cite{Mc02} or
by invoking a high neutronization during the SN explosion, which would
lead to the direct synthesis of iron. In this latter case, however, no
contribution from the SN should be observed in the afterglow,
contrary to recent observations~\cite{Bl02}.

\subsection{The spectrum of GRB~011211}

The spectral features detected in the spectrum of GRB~011211 may
represent a cornerstone in our understanding of the line production
mechanism. In fact, the detection of large EW $K_\alpha$ lines from
elements such as Mg, Si, S, Ar and Ca~\cite{Re02}, can be explained,
within the framework of reflection models, only if the continuum
radiation that illuminates the reflecting material is hidden to the
observer~\cite{La02}. In fact, if a reflected spectrum is added to the
incident one, the EW of S lines cannot be larger than $\sim 100$~eV,
while Reeves et al.~\cite{Re02} detected a S line with
$EW=480\pm140$~eV, which can be explained only if the emitted line is
not diluted into the incident continuum radiation.

Such a condition can be achieved in a contrived geometry or, more
simply, can be due to light travel effects, if the reflecting material
lies at a certain distance from the burst explosion site away from the
line of sight~\cite{La99}. The observer at infinity will in fact
receive the undeflected ionizing continuum at a time $t_0$ and the
reflected component at a time $t_0+R/c(1-\cos\theta)\gg{t}_0$ where
$\theta$ is the angle between the line of sight and the line
connecting the burst explosion site to the reflecting material.

The spectrum of GRB~011211 strongly suggests, therefore, that lines
are produced in a GD scenario. It should be mentioned, however, that
the line detection, albeit confirmed by the authors~\cite{Re02a}, has
been challenged on data analysis~\cite{Bo02} and on
statistical~\cite{Ru02} grounds, and all the conclusion above are
therefore subject to revision, should the detection be proved false.

\section{Line and burst energetics}

Besides being a powerful probe of the condition of the close
environment of GRBs, emission lines provide us with an isotropic
emission that can be used to constrain the total energetics of the
event~\cite{Gh02}.  The total energy of the burst as derived from the
line emission can be written as:
\begin{equation}
E=E_\mathrm{line,~iso} \quad \epsilon_\mathrm{line}^{-1} \quad
\epsilon_\mathrm{spex}^{-1} \quad \epsilon_\mathrm{eff}^{-1}
\label{eq:etot}
\end{equation}
where $E_\mathrm{line,~iso}$ is the isotropic equivalent energy
observed in line photons and $\epsilon_\mathrm{line}^{-1}$ is the
efficiency of line production, i.e. the ratio of the line to incident
continuum luminosity. $\epsilon_\mathrm{spex}^{-1}$ is the fraction of
the burst spectrum in the X-ray band and $\epsilon_\mathrm{eff}^{-1}$
is the ratio of total energy to the energy in photons.

\subsection{Line energetics}

In order to obtain a measure of the total energy from
Eq.~\ref{eq:etot} one has to estimate the total energy that is
observed in the form of line photons. Besides the uncertainty on the
line flux itself, the main problem is to evaluate the duration of the
line emission. In fact, the observations start usually some time after
the GRB, so that it is not possible to know the line flux at very
early times. In addition, in several cases it is not possible to set a
time for its turn-off time. In Tab.~\ref{tab1} we report the line
energetics assuming that the line had a constant flux from the time of
the GRB explosion up to the longer time in which it was detected,
either the last observation or the moment in which it was observed to
fade. See Ghisellini et al.~\cite{Gh02} for a more complete
discussion.

\subsection{Line efficiency}

In this section we evaluate the efficiency in the production of line
photons by the reflection mechanism. Even though an accurate result
can be obtained only with a numerical treatment~\cite{La02, Gh02}, we
discuss here an approximate treatment that gives correct estimates. We
concentrate here on iron lines, referring the reader to the literature
for a more complete description~\cite{Gh02}.

The line emission efficiency can be estimated as the ratio of the
energy in the photons absorbed by iron atoms in the layer of material
with $\tau_\mathrm{Fe}=1$ to the total energy continuum photons,
corrected for the fact that the energy of the line photon is lower
than that of the photoionizing one. This yields:
\begin{equation}
\epsilon_\mathrm{line} \sim
{{h\nu_\mathrm{line} \int_{\nu_\mathrm{ph}}^{\nu_M} {{{L(\nu)}\over{h\nu}}
\left({{\nu}\over{\nu_\mathrm{ph}}}\right)^{-3} \,d\nu}} \over
{\int_{\nu_m}^{\nu_M} L(\nu)\,d\nu}}
\label{eq:eli}
\end{equation}

\begin{figure}[htb]
\begin{center}
\parbox{0.48\textwidth}{\epsfig{file=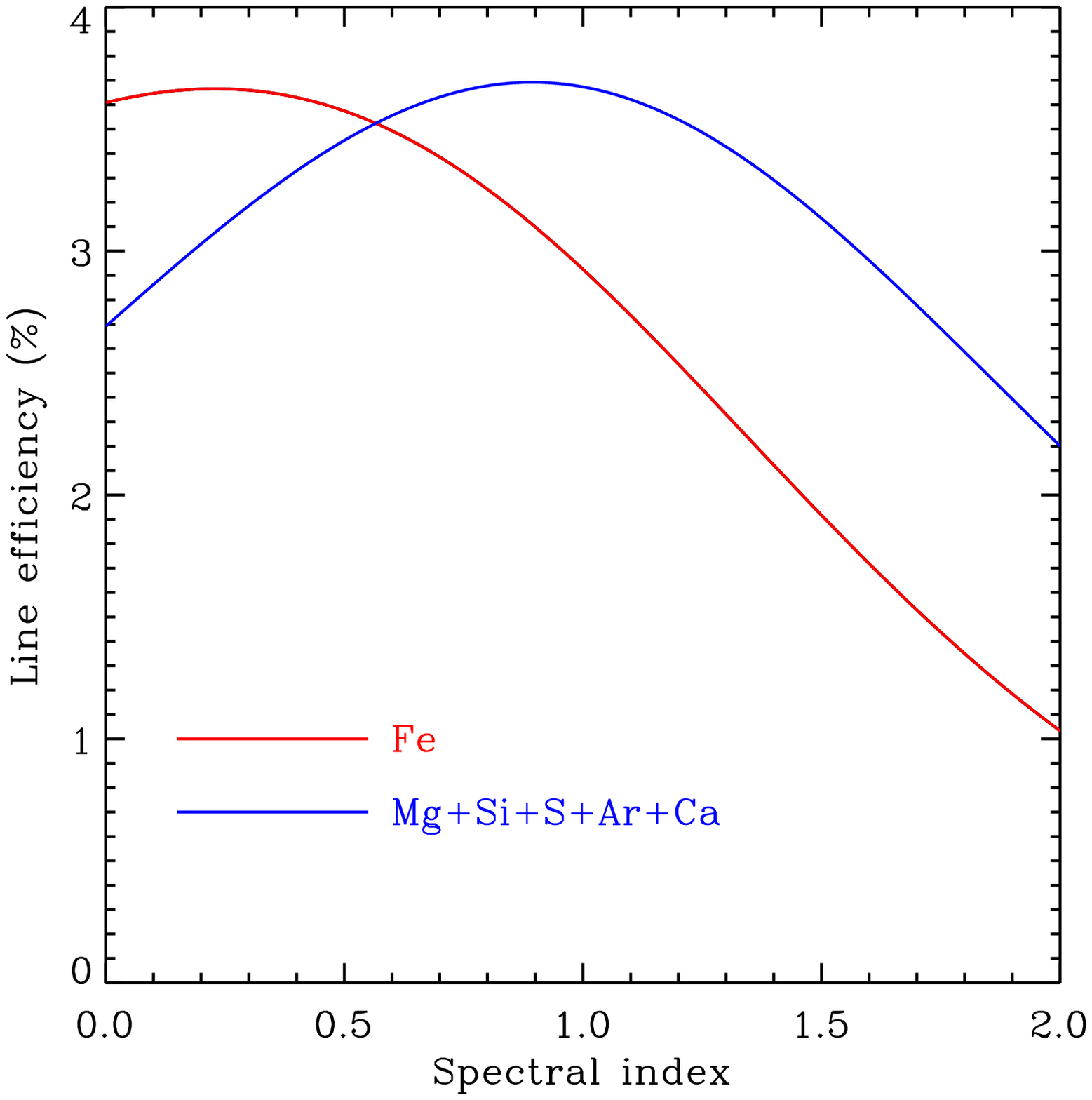,width=0.47\textwidth}
\caption{{Line production efficiency in reflection models for iron 
(red line) and lighter elements as a function of the spectral index of
the ionizing continuum. The lines show the result of the approximate
Eq.~\ref{eq:eli}.}
\label{fig:eli}}
\vspace{1truecm}}
\hspace{0.01\textwidth}
\parbox{0.48\textwidth}{\epsfig{file=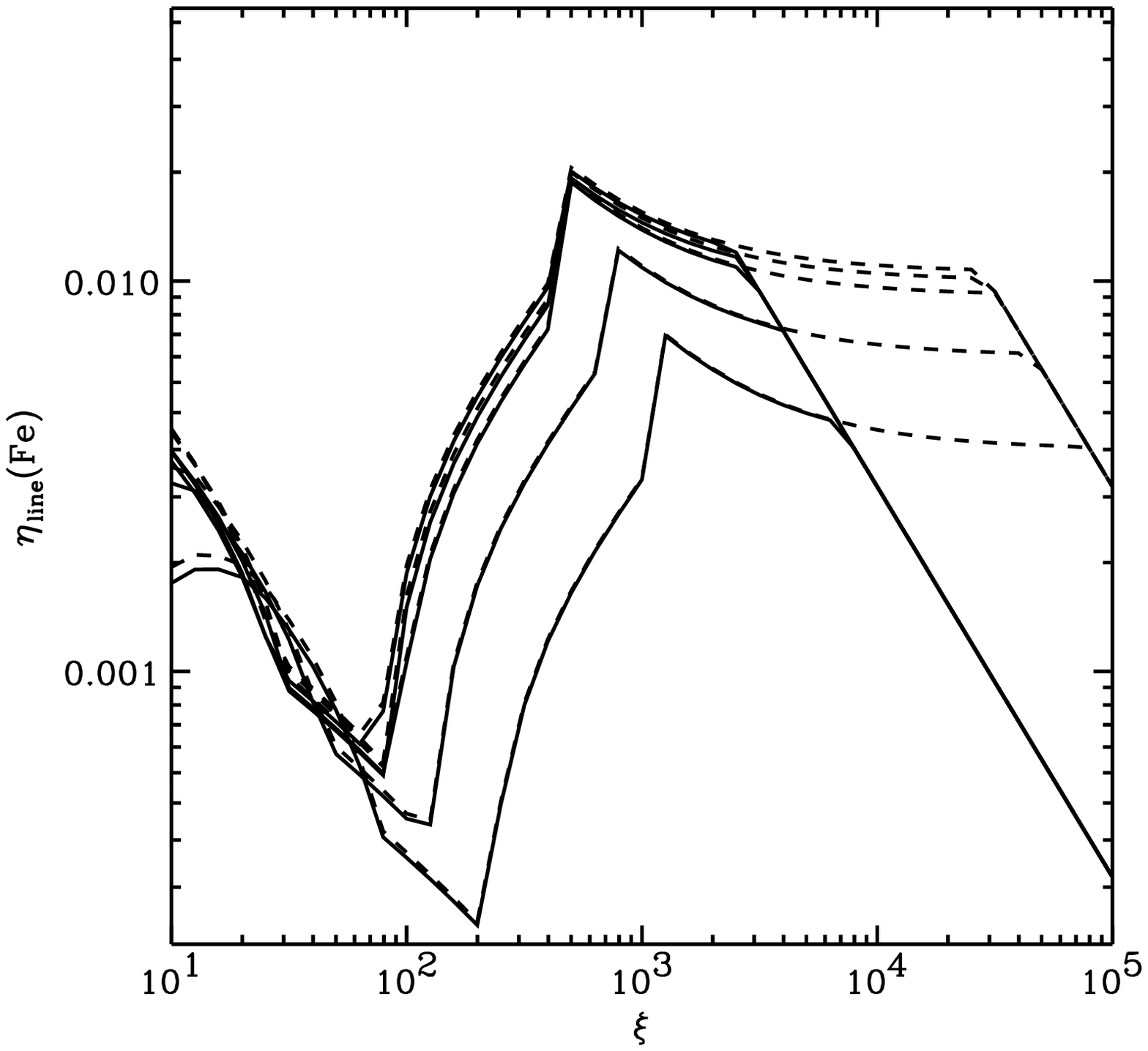,width=0.47\textwidth}
\caption{{Line production efficiency in reflection models for iron from 
the numerical computations of Lazzati et al.~\cite{La02}. The
efficiency is plotted against the ionization parameter, while
different lines are relevant to different spectral indices and
metallicities (see the above reference for more details).}
\label{fig:eli2}}}
\end{center}
\end{figure}

The integral of Eq.~\ref{eq:eli} can be solved analytically for a
power-law input spectrum, yielding the efficiency plotted in
Fig.~\ref{fig:eli} as a function of the power-law index $\alpha$
[$L(\nu)\propto\nu^{-\alpha}$]. The approximation gives the maximum
allowed efficiency, since it neglects the effect of the ionization
parameter and of Auger autoionization. A numerical
treatment~\cite{La02} returns nevertheless consistent results within a
factor of two (Fig~\ref{fig:eli2}), showing that this parameter is
safely constrained to few per cent.

\subsection{Bolometric correction}

The bolometric correction to transform the X-ray energy into total
radiative energy can be computed by assuming a Band spectrum with
average values of the parameters (the two spectral slopes and the peak
energy). This yields $\epsilon_\mathrm{spex}\approx0.03\div0.1$.

\subsection{Radiative efficiency}

The ratio of the energy in radiation to the total energy is the most
uncertain and cannot be constrained. It should be remarked, however,
that in order to compare the total energy inferred from line emission
to the one estimated by Frail et al.~\cite{Fr01}, this correction is
not necessary.

\section{The total energy from lines}

Before discussing how the total energy derived from line emission
compares to other measurements it is worth discussing what we mean by
the energetics of the GRB phenomenon. There are in fact at least three
possible definitions of the total energy of the bursts.

The first possible definition is the {\it real total energy involved
in the explosion}. This includes the energy that is released in
gravitational waves, neutrinos, the possibly associated SN explosion
and the burst itself. It may be of the order of $\sim10^{54}$~erg in
most of the progenitor models.  Then, it is possible to define a {\it
usable energy for the GRB}, which includes all the forms of energy
that are more strictly related to the GRB phenomenology and/or can
contribute emission during the burst and afterglow phases of the GRB
explosion. This energy includes all the fireball energy, any precursor
activity in photons or kinetic energy, any delayed output from the
inner engine and energy stored in the jet cocoon (as discussed
above). Is is clearly smaller than the {\it real total energy involved
in the explosion}.  Finally, what is more easy to measure is the {\it
fireball energy}, and in particular the energy released in photons by
the fireball. This is quite easy to constrain since we can measure the
burst fluence and we have some tools to correct this energy input for
any non-isotropy of the emission. This energy is clearly the smallest
of the three.

Now, if we compute the total energy from Eq.~\ref{eq:etot}, applying
the fiducial values discussed above for the efficiencies involved, we
obtain a rather large number (the data of GRB~991216 have been used as
an example case):
\begin{equation}
E=3\times10^{52} \, \left({{0.03}\over{\epsilon_\mathrm{line}}}\right)
\, \left({{0.03}\over{\epsilon_\mathrm{spex}}}\right)\,
\epsilon_\mathrm{eff}^{-1} \quad \mathrm{erg}
\end{equation}
to be compared with the value of $E\sim5\times10^{50}$ measured by
correcting the $\gamma$-ray isotropic energy times the solid angle of
the jet, as derived from afterglow observations~\cite{Fr01, PK02}.

The discrepancy is quite significant and calls for an
explanation. There are two possible answers. First, the determination
of the opening angle of the jet from the break time is prone to
systematic uncertainties, related to the density of the ISM and to a
possible missing constant into the equations (see e.g. the different
approach of Rhoads~\cite{R99} and Halpern et al.~\cite{Sa99}). It is
therefore possible that a systematic underestimate of the ISM density
caused an underestimate of the opening angles and therefore of the
total energy. Since, however, the ISM density comes into the equations
with a very small power, this correction may ameliorate the discrepancy
but it is unlikely to solve it completely.

Alternatively, it may be possible that the energy inferred from line
emission is related to the {\it usable energy for the GRB} defined
above rather than to the {\it fireball energy}. In this case, since
the three energies discussed above define a nested sequence, there
would not be any problem. However, if the {\it fireball energy} is
indeed a very small fraction of the {\it usable energy for the GRB},
which is a very small fraction of the {\it real total energy involved
in the explosion}, it is rather surprising that the {\it fireball
energy} seems to be a remarkably well defined quantity~\cite{Fr01,
PK02}.

\bigskip
I am grateful to Martin Rees, Gabriele Ghisellini, Elena Rossi and
Enrico Ramirez-Ruiz for the collaboration, the discussions and all the
work done together.

\def\Discussion{
\setlength{\parskip}{0.3cm}\setlength{\parindent}{0.0cm}
     \bigskip\bigskip {\Large {\bf Discussion}} \bigskip}
\def\speaker#1{{\bf #1:}\ }
\def\endDiscussion{}

\Discussion

\speaker{S. Kulkarni}  I am concerned that your adopted efficiency to 
convert the X-ray flux to line production is optimistic. Specifically,
for 011211, with a narrow jet opening combined with high Lorentz
factor would result in a photon beam which is quite narrow, a few
degrees across.  Does this not mean that you have to arrange your
reflection surface in a very contrived way?

\speaker{D. Lazzati} What I am saying is that the lines require a 
total energy larger than what we obtain if we correct the $gamma$-ray
observations for the beaming angles reported in the literature. If we
relax this requirement, we can envisage a geometry in which a sizable
fraction of the X-ray continuum is reprocessed. For example the walls
of the reflecting funnel can be concave, or the reprocessing material
may be in the form of clumps or filaments in a smaller version of the
Crab remnant.

\endDiscussion
 
\end{document}